\documentclass[%
twocolumn,superscriptaddress,
 amsmath,amssymb,
 aps,
prc,
]{revtex4-1}
\usepackage{graphicx}% Include figure files
\usepackage{dcolumn}
\begin{document}

% Use the \preprint command to place your local institutional report
% number in the upper righthand corner of the title page in preprint mode.
% Multiple \preprint commands are allowed.
% Use the 'preprintnumbers' class option to override journal defaults
% to display numbers if necessary
%\preprint{}

%Title of paper
\title{Energy partition between splitting fission fragments}

% repeat the \author .. \affiliation  etc. as needed
% \email, \thanks, \homepage, \altaffiliation all apply to the current
% author. Explanatory text should go in the []'s, actual e-mail
% address or url should go in the {}'s for \email and \homepage.
% Please use the appropriate macro foreach each type of information

% \affiliation command applies to all authors since the last
% \affiliation command. The \affiliation command should follow the
% other information
% \affiliation can be followed by \email, \homepage, \thanks as well.
\author{H.Y. Shang}
\affiliation{
State Key Laboratory of Nuclear Physics and Technology, School of Physics,
Peking University, Beijing 100871, China
}
\author{Y. Qiang}
\affiliation{
State Key Laboratory of Nuclear Physics and Technology, School of Physics,
Peking University, Beijing 100871, China
}
\author{J.C. Pei}\email{peij@pku.edu.cn}
\affiliation{
State Key Laboratory of Nuclear Physics and Technology, School of Physics,
Peking University, Beijing 100871, China
}
\affiliation{
Southern Center for Nuclear-Science Theory (SCNT), Institute of Modern Physics, Chinese Academy of Sciences, Huizhou 516000,  China
}

%Collaboration name if desired (requires use of superscriptaddress
%option in \documentclass). \noaffiliation is required (may also be
%used with the \author command).
%\collaboration can be followed by \email, \homepage, \thanks as well.
%\collaboration{}
%\noaffiliation

\date{\today}

\begin{abstract}
% insert abstract here
From the microscopic view, the energy partition between two fission fragments are associated with
the splitting of wave functions of an entangled fissioning system,  in contrast to most fission models
using an explicit statistical partition of excitation energies by invoking level densities of fragments.
The dynamical fission evolution is described within the time-dependent Hartree-Fock+BCS framework.
Excitation energies of isotopic fission fragments are obtained with the particle-number projection method
after the dynamical splitting of $^{238}$U.
The resulting excitation energies of light and heavy fragments illustrate the appearance of sawtooth structures.
We find that the paring strengths have significant influences on the partition of excitation energies.
Furthermore,  excitation energies of isotopic fragments increase with increasing neutron numbers,
suppressing the production of neutron-rich beams in rare-isotope beam facilities.

\end{abstract}

% insert suggested keywords - APS authors don't need to do this
%\keywords{}

%\maketitle must follow title, authors, abstract, and keywords
\maketitle

% body of paper here - Use proper section commands
% References should be done using the \cite, \Ref.~and \label commands
% \section{The NN potential and local projection}
\emph{Introduction.}---
In the final stage of nuclear fission, the majority of nuclear energy is released through the enormous
kinetic energies of splitting fission fragments.
However, the remain considerable part of nuclear energy is stored as excitation energies
of primary fission fragments. Consequently, the de-excitation fission fragments is realized  by neutron emissions, $\gamma$ radiations, and then $\beta$-decays~\cite{future}.
Therefore, the energy partition between two fragments plays an indispensable role in
determining multiple post-fission observables and their correlations.

The excitations of fragments are caused by the dissipative motion and shape distortions.
The dissipation can play a significant role even in the final splitting stage~\cite{qy2}.
The shape distortion and shell effects of fragments are important in the energy partition~\cite{albertsson}.
The non-equilibrium non-adiabatic fission dynamics,  shell effects, dynamical pairing correlations,
shapes of primary fragments,
and energy dependencies can be naturally described by the microscopic time-dependent density functional theory (TD-DFT)~\cite{koonin,negele,tddft,simenel2020,stevenson,bulgacprl,bulgacprc,scampsnature,formation,ayikprl,qy1,qy2,shiyue}.
It has been pointed out that the energy partition is later than particle partition~\cite{bulgac-energy}.
In previous fission studies of $^{240}$Pu, the light fragments acquire more excitation energies than that of
heavy fragments~\cite{qy1}. This is consistent with observations that light fragments emit more neutrons
compared to heavy fragments from the fission of actinide nuclei. With increasing excitation energies, the difference in excitation energies between light and heavy fragments
are reduced~\cite{qy1}.

Experimentally, the average number of  neutrons, i.e., neutron multiplicities,  emitted from fission fragments shows
the puzzling sawtooth structures in dependence of fragment masses~\cite{neutron1,neutron2}.
This provides a unique chance to understand the energy sharing between two fission fragments.
Conventionally the excitation energy sharing between fission fragments are described
as at the statistical equilibrium by invoking the different level densities of fragments~\cite{albertsson,schmidt,chen,fujio,mora}.
It has been pointed out that shape-dependent level densities can better describe the partition of excitation energies~\cite{albertsson}.
Within this approach, the slopes of the sawtooth structures are slightly underestimated~\cite{albertsson}.
Usually different temperatures in light and heavy fragments have to be adopted to reproduce neutron multiplicities~\cite{chen,fujio}.
The sawtooth structures are also shown in the distributions of neutron excess in dependence of fragment-charge number~\cite{nzratio2,nzratio3}
and angular momentum in dependence of fragment masses~\cite{wilson}.

Recently, we proposed that the quantum entanglement is crucial for the appearance of sawtooth structures
in the distributions of excitation energies of fragments and subsequently neutron multiplicities~\cite{qy3}.
The sharing of particles between two fission fragments obtained by double particle number projection (PNP)
shows a considerable spreading width~\cite{qy3,schunckproj}.
The associated energy partition due to the superposition of different particle numbers can be obtained by the PNP method~\cite{qy3}.
The distribution of fragment yields has been studied by PNP within the framework of time-dependent generator-coordinate-method
and TD-DFT~\cite{schunckproj,schunckp2,bulgacproj}.
The PNP method has also been used to study heavy ion reactions~\cite{sekizawa,godbey2020}.
In most fission models, the quantum correlations or entanglement between two fission fragments have not been taken into account.
It is timely to study energy  partition between fission fragments considering quantum entanglement between two fission fragments,
in which the entanglement is persistent even two fragment are well separated~\cite{qy3}.

In this work, we studied the partition of excitation energies of isotopic fission fragments of $^{238}$U, which is relevant for the production of radioactive beams from the fission products after prompt neutron evaporations.
For example, the medium-mass neutron-rich beams are mainly produced by the frisson of $^{238}$U in new-generation rare isotope beam facilities such as FRIB~\cite{frib}, RIBF~\cite{ribf} and HIAF~\cite{hiaf}.
Three new isotopes have been produced in fission reactions of $^{238}$U beam in the carbon target in newly operated FRIB~\cite{frib}.
Furthermore, the proposal of photo-fission of $^{238}$U driven by a high-power e-LINAC with  a convertor target is promising to produce neutron-rich beams~\cite{proposal}.

\emph{Methods}---
The time-dependent Hartree-Fock+BCS (TD-BCS) approach is used
to describe the dynamical fission evolution beyond the saddle
point.
 The TD-BCS equations can be derived by using the
BCS basis or canonical basis in the time-dependent Hartree-Fock-Bogoliubov method~\cite{TDBCS1}.
In our previous work, TD-BCS has been extended to
study fission dynamics of compound nuclei~\cite{qy1}.
The initial configuration of the fissioning nucleus
is obtained by deformation-constrained Hartree-Fock+BCS
calculations. We employed  constrained calculations in terms of quadrupole-octupole
deformations $(\beta_2, \beta_3)$.
The initial deformation is adopted as $(\beta_2, \beta_3)$=(2.4, 1.4) for the fission evolution.
For nuclear interactions, SkM$^{*}$~\cite{skm} and UNEDF1~\cite{unedf} forces have been adopted,
which have been widely used for calculations of nuclear fission barriers.
The mixing-type pairing interaction~\cite{mix-pair} is adopted with strengths $V_p$=475 MeV and $V_n$=420 MeV for SkM$^{*}$ force,
and $V_p$=415 MeV and $V_n$=375 MeV for UNEDF1 force.
The dynamical evolution is performed with the time-dependent Hartree-Fock (TDHF)
solver Sky3D~\cite{sky3d} with our modifications of TD-BCS.
The initial configurations are obtained using the SkyAX solver~\cite{skyax} to
interface with Sky3D.

Based on TD-BCS solutions, the particle numbers of fragments
and the fissioning nucleus are not well defined.
The particle numbers in two fragments are in a superposition state.
 The double PNP
on the total space and a partial space is applied to determine
the particle numbers of two complementary fragments.

The double projection operator is written as
\begin{equation}
\begin{aligned}
    \hat{P}^q\left(N_T^q, N_P^q\right)= & \frac{1}{4 \pi^2} \iint d \theta_T d \theta_P \\
    & \mathrm{e}^{\mathrm{i} \theta_T\left(\hat{N}_T^q-N_T^q\right)} \mathrm{e}^{\mathrm{i} \theta_P\left(\hat{N}_P^q-N_P^q\right)},
\end{aligned}
\end{equation}
where $q$ denotes neutron or proton, $T/P$ denotes the projection on the
total space or a partial space and $\hat{N}^q_P$ denotes the particle number
operator. Note that the particle number operator is
$\hat{N}^q_P=\int d \boldsymbol{r} \hat{C}^{\dagger}(\boldsymbol{r}) \hat{C}(\boldsymbol{r})\Theta(\boldsymbol{r})$,
where $\Theta(\boldsymbol{r})$ is a mask function to obtain an exclusive partial space.
The projected states with particle numbers deviating from the average number up to 8 particles are calculated.

The PNP on the wave functions of each fragment  leads to a two-dimensional distribution of fragments in terms of ($Z$, $N$), in which
the formation probability of each fragment  is the expectation value of
$\langle\Psi| \hat{P}^\text{n}\left(N_T, N_P\right) \hat{P}^\text{p}\left(Z_T, Z_P\right)|\Psi\rangle$.
The projected binding energy of each fragment is obtained by
\begin{equation}
    E_{\text {proj}}=\frac{\langle\Psi| \hat{H} \hat{P}^\text{n}\left(N_T, N_P\right)
    \hat{P}^\text{p}\left(Z_T, Z_P\right)|\Psi\rangle}
    {\langle\Psi| \hat{P}^\text{n}\left(N_T, N_P\right) \hat{P}^\text{p}
    \left(Z_T, Z_P\right)|\Psi\rangle}, \label{eq:Eproj}
\end{equation}
which is actually calculated as~\cite{egido,sheikh}
\begin{equation}
    \begin{aligned}
        E_{\text{proj}} & =\int d \theta_\text{n} \int d \theta_\text{p} Y_{\theta_\text{n}} Y_{\theta_\text{p}} \operatorname{Tr}\left\{t\left(\rho_{\theta_\text{n}}^\text{n}+\rho_{\theta_\text{p}}^\text{p}\right)\right. \\
        & +\frac{1}{2}\left(\Gamma_{\theta_\text{n}}^{\text{n}\text{n}} \rho_{\theta_\text{n}}^\text{n}+\Gamma_{\theta_\text{p}}^{\text{p}\text{p}} \rho_{\theta_\text{p}}^\text{p}+\Gamma_{\theta_\text{p}}^{\text{n}\text{p}} \rho_{\theta_\text{n}}^\text{n}+\Gamma_{\theta_\text{n}}^{\text{pn}} \rho_{\theta_\text{p}}^\text{p}\right. \\
        & \left.\left.-\Delta_{\theta_\text{n}}^\text{n} \bar{\kappa}_{\theta_\text{n}}^{\text{n*}}-\Delta_{\theta_\text{p}}^\text{p} \bar{\kappa}_{\theta_\text{p}}^{\text{p*}}\right)\right\} \\
        Y_{\theta_q} & =\langle\Psi| \mathrm{e}^{\mathrm{i} \theta_T\left(\hat{N}_T^q-N_T^q\right)} \mathrm{e}^{\mathrm{i} \theta_P\left(\hat{N}_P^q-N_P^q\right)}|\Psi\rangle /\langle\Psi| \hat{P}^q|\Psi\rangle,
        \end{aligned}
        \label{eq:Eproj2}
\end{equation}
where $\rho_\theta$, $\kappa_\theta$ are transition densities.
The excitation energy of each fragment is obtained by the subtraction between the projected binding
energy in the splitting process and the ground state energy.
This method has been applied to calculate excitation energies of products in
multi-nucleon transfer reactions~\cite{sekizawa}.
In practical calculations, a series of transition densities
like the current density, spin-orbit density have
to be calculated at each $\theta$. The calculations are
very costly because the proton-neutron mixing
terms involve fourfold integrations.
The calculations could be problematic when the denominator in Eq.~\eqref{eq:Eproj} is tinny,
and a cutoff at $5\times10^{-4}$ is applied.

\begin{figure}
    \includegraphics[width=\columnwidth]{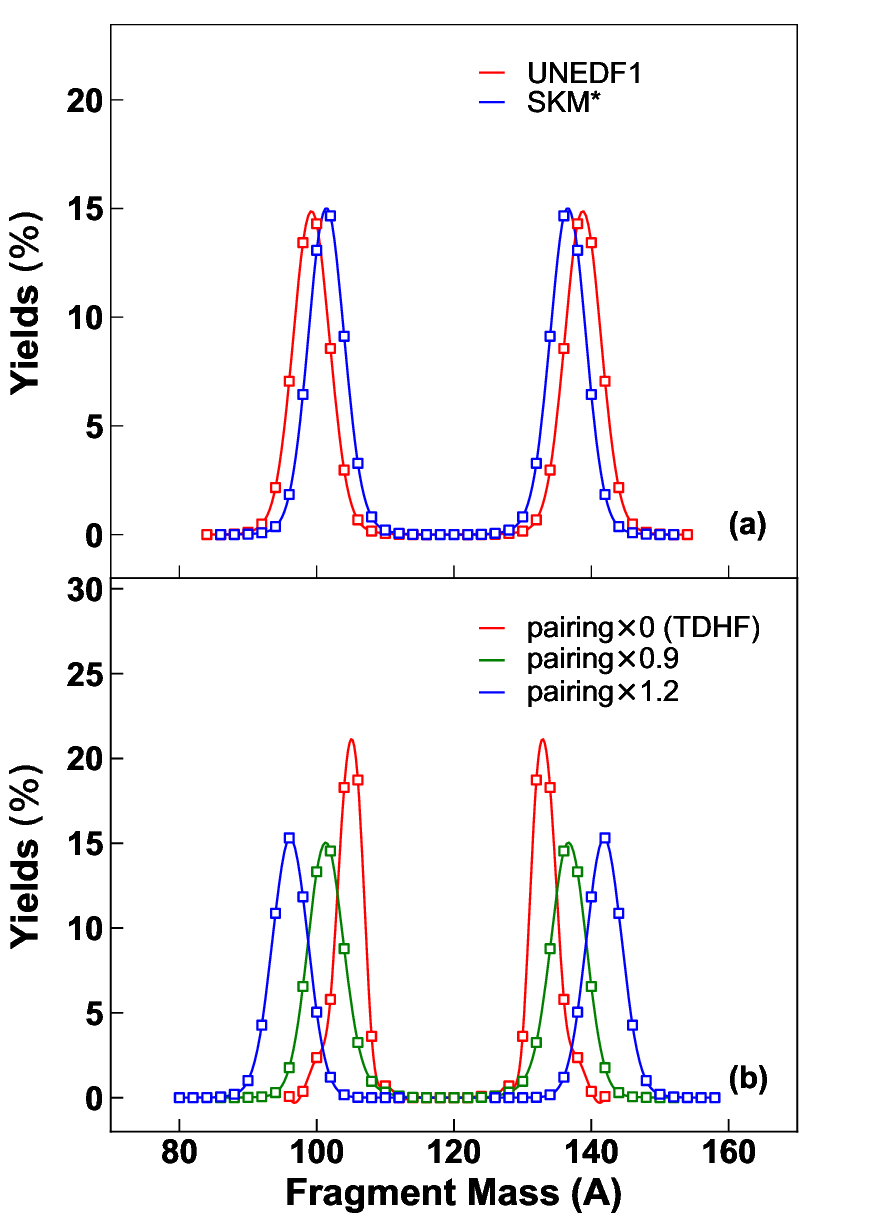}\\
    \caption{%\label{fig:VT_NLO}
Fission yields of $^{238}$U based on TD-BCS+PNP calculations  of the splitting fission fragments.
(a) Results obtained with SkM$^{*}$ and UNEDF1 forces, respectively. (b) Results obtained with SkM$^{*}$
force and varying pairing strengths corresponding to factors of 0.0 (TDHF), 0.9 and 1.2, respectively.
    } \label{Fig1}
\end{figure}

%\subsection{fission yields with PNP}

\begin{figure}
    \includegraphics[width=\columnwidth]{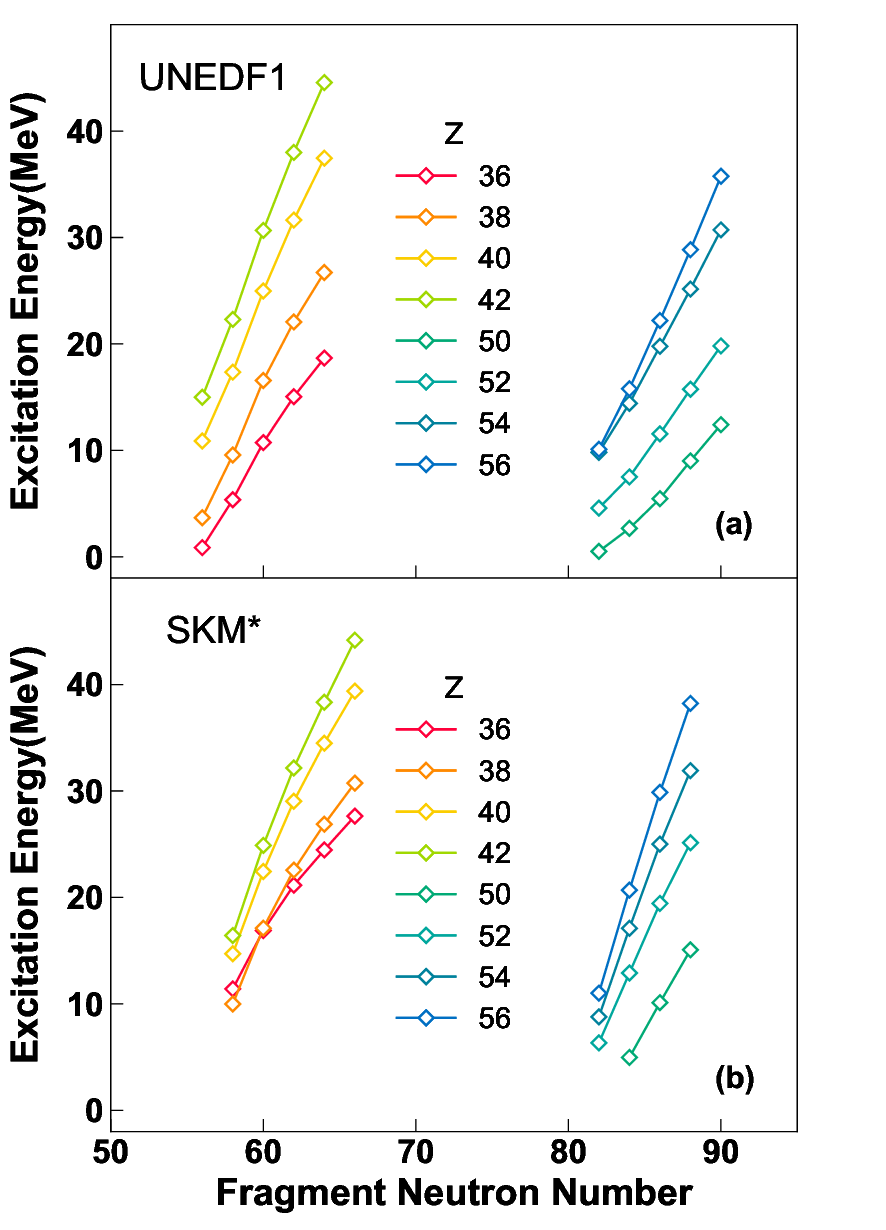}
    \caption{%\label{fig:VT_NLO}
Excitation energies of isotopic fission fragments of $^{238}$U after PNP on the splitting fission fragments.
(a) Results obtained with SkM$^{*}$ force;  (b) Results obtained with UNEDF1 forces.
    }
    \label{Fig2}
\end{figure}

\emph{Results.}---
Firstly the distributions of fission yields of $^{238}$U after PNP on the splitting frisson event
are obtained. Fig.\ref{Fig1}(a) shows the distributions of projected fission yields as a function of fragment masses,
calculated with SkM* and UNEDF1 forces, respectively.
It can be seen that the peak is around A=136 with SkM$^{*}$ force, but the peak is around A=138  with UNEDF1 force.
The peak widths are similar and the half-widths are about 8 number of particles.
It is known that UNEDF1 results in a slightly lower fission barriers than that of SkM$^{*}$~\cite{unedf}. This could be
a reason that the fission yield peak from UNEDF1 calculations is slightly more asymmetric.
Correspondingly, its total kinetic energy (TKE) is smaller
with a longer scission neck. The resulting TKE of SkM* calculations is 168.9 MeV while TKE of UNEDF1 calculations is about 159.5 MeV.
Note that the average experimental TKE from photofission of $^{238}$U is around 170 MeV~\cite{tke}.

Pairing correlations are important in descriptions of fission probabilities~\cite{witek} and dynamical fission evolutions~\cite{bulgacprl,qy1,xiaobao}.
To study the role of pairing correlations in dynamical calculations of particle partition  between fission fragments,
the distributions of fission yields after PNP are displayed with varying pairing strengths, as shown in Fig.\ref{Fig1}(b).
It can be seen that the peak locations are shifted to more asymmetric fission modes with increasing pairing strengths.
With TDHF calculations without pairing correlations, the peak location is around A=132 and is close to the asymmetric S1 fission channel~\cite{brosa, modes}.
This situation is similar to our previous studies of the fission of $^{240}$Pu~\cite{qy1}.
The peak locations from calculations with pairing strengths reduced by a factor of 0.9 are close to the original results.
However, the peak location is shifted to A=142 if the pairing strength increased by a factor of 1.2.
It has been pointed out that a larger pairing strength  results in a longer scission neck~\cite{qy2}, which could be related to more asymmetric fission yields.
Within TDHF, the width from PNP is obviously smaller than that of TD-BCS+PNP calculations.
This implies that the spreading width of fission yields can be enhanced by including many-body correlations.
It is known that the width of S1 channel is indeed narrower than that of S2 channel~\cite{brosa}.
With increasing pairing strengths, the scission neck becomes longer and the resulting TKE become smaller.
The TKE corresponding to pairing factors of 0.0, 0.9 and 1.2 are 174.2, 169.6 and 160.3 MeV, respectively.

%\subsection{excitation energies of fragments}
It is known that average excitation energies of fission fragments can be obtained by TD-BCS calculations without PNP.
Within TD-BCS, the light fragment has higher excitation energies than that of the heavy fragment at low energy fission~\cite{qy1}.
In this work, we are interested to study the distribution of excitation energies of all fission fragments with PNP.
This is related to the intriguing sawtooth structures of neutron multiplicities.

Fig.\ref{Fig2} shows the resulting excitation energies of different isotopic fragments.
It can be seen that UNEDF1 results are similar to that of SkM* results.
The results of isotopes around Z=46, i.e., the symmetric fission channel, having very small yields, are not shown.
Usually the average neutron multiplicities are illustrated in terms of fragment masses.
In our calculations, the detailed results shows that the distributions of excitation energies of isotopic fragments have positive slopes.
This actually explains the origin of sawtooth structures.
Generally light fragments have higher excitation energies than that of heavy fragments.
However, this is not the case for two complementary fragments.
For the same isotopic fragments, the heavier fragments have higher excitation energies.
It has to be noted that, to obtain realistic two-dimensional distributions of fission yields and excitation
energies, fluctuation effects in the fission process should be taken into account,
which can
significantly alleviate the slopes of sawtooth structures~\cite{qy3}.

In Fig.\ref{Fig2}, excitation energies of isotopic fragments increase as the neutron number increases, which has
significant implications. Currently the rare isotope beam facilities~\cite{frib,ribf} mainly
rely on the acceleration of fission fragments from $^{238}$U.
In particular, the Coulomb excitation induced fission of $^{238}$U with a high-$Z$ target has advantages for production of neutron-rich beams~\cite{highz}.
The reliable estimation of beam intensities is a practical issue.
The beam intensity calculations are conventionally based on the code LISE$^{++}$~\cite{lise}, which relies on empirical fission yields.
It is difficult to simultaneously describe intensities of  light and heavy fragments~\cite{ribf}.
In our calculations, the heavier isotopes have higher excitation energies, resulting
more neutron evaporations. This means that the production of neutron-rich rare isotope beams would be
suppressed.
The partition of excitation energies would be changed at high energies when sawtooth structures are also washed out~\cite{washout}.

\begin{figure}
    \includegraphics[width=\columnwidth]{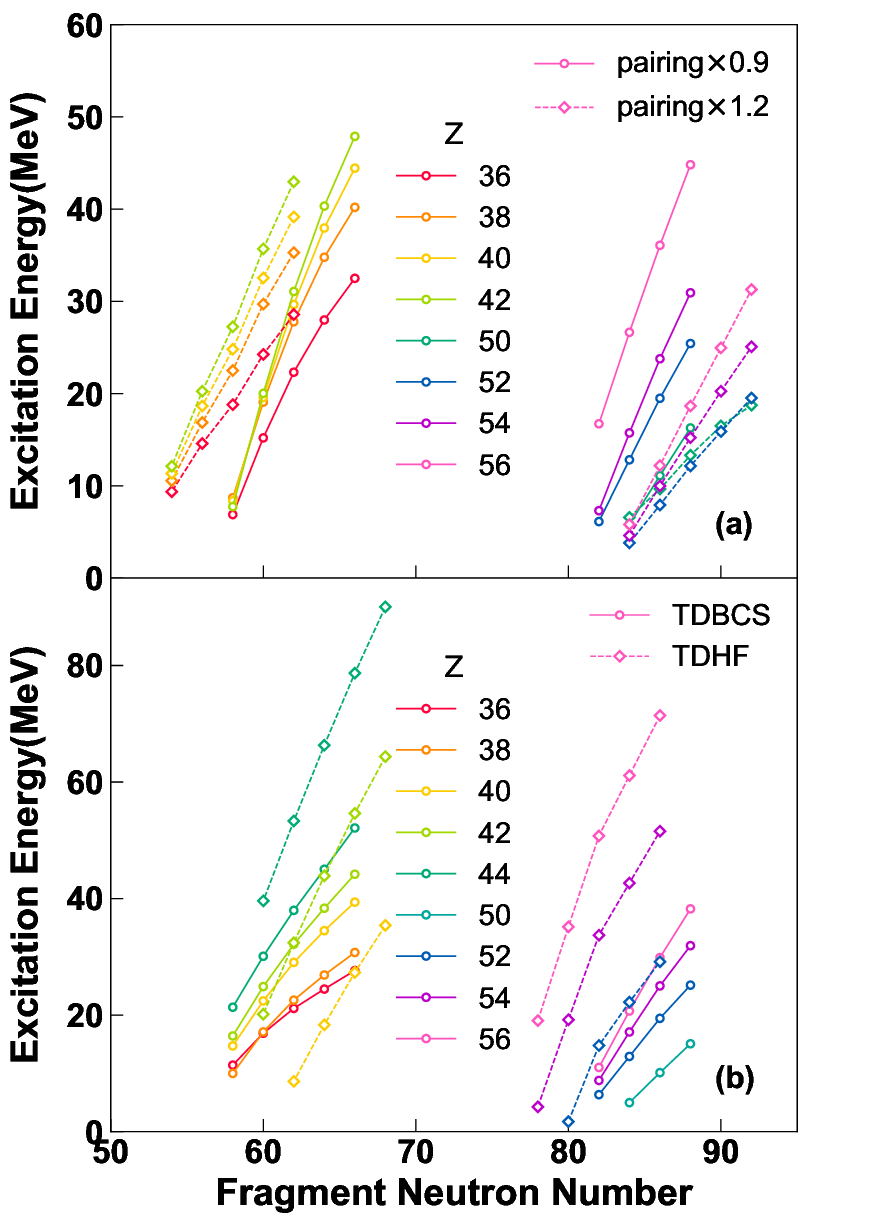}
    \caption{%\label{fig:VT_NLO}
 Excitation energies of isotopic fission fragments from $^{238}$U after PNP on the splitting fission fragments.
 (a) Results obtained with pairing strengths varied by a factor of 0.9 and 1.2 respectively.
 (b) Results obtained with TDHF (zero pairing) and TD-BCS methods.
    }\label{Fig3}
\end{figure}

Fig.\ref{Fig3} shows the excitation energies of fission fragments with varying pairing strengths.
With increasing pairing strengths, the fission yield peaks would be slightly more asymmetric, as shown in Fig.\ref{Fig1}(b).
It can be seen that the maxima excitation energies of fragments are generally reduced with increasing pairing strengths.
In particular, excitation energies of heavy fragments except for the $Z$=50 shell decrease significantly with increasing pairings.
In addition, the slopes of excitation energies of light fragments are reduced with increasing pairing strengths.
The slopes are overestimated within our approach, which can be alleviated by fluctuation effects~\cite{qy3}.
Fig.\ref{Fig3}(b) shows the excitation energies from TD-BCS+PNP and TDHF+PNP calculations.
It can be seen that the excitation energies and its slopes from TDHF+PNP are too large,  corresponding to the narrow S1 channel.
Our results indicate that many-body correlations in addition to pairing correlations
might be useful to obtain reasonable excitation energies of fragments as well as to alleviate the associated slopes.
In this respect, fluctuations can be seen as an effective treatment of high-order correlations.
Nevertheless, to quantitatively reproduce the distributions of fission yields and neutron multiplicities microscopically is
beyond the present work.

\emph{Summary.}---
The excitation energies of isotopic fission fragments from $^{238}$U are calculated with the microscopic TD-BCS plus PNP method
for deeper understandings of nuclear fission.
This is different from most fission models that use the explicit statistical partition of excitation energies between fragments.
The dependencies of the energy partition on different Skyrme forces and varying pairing strengths are studied.
With increasing pairing strengths, the peak of fission yields shifts to a slightly more asymmetric fission channel.
Within TDHF calculations without pairings, the width of fission yields is rather narrow and its peak is close to
the S1 fission channel.
The obtained excitation energies of isotopic fission fragments explain the origin of sawtooth structures.
Furthermore, pairing correlations play a significant role in the partition of energies between fragments.
The slope of excitation energies of light fragments would decrease with increasing pairing strengths.
For heavy fragments, the excitation energies of heavy fragments except for the $Z$=50 shell would decrease significantly with
increasing pairing correlations.
The excitation energies based on TDHF+PNP are too large and the associated slopes are very steep.
Our results indicate that many-body correlations or fluctuations are essential to obtain reasonable excitation energies.
It has to be pointed out that the excitation energy partition and consequently neutron evaporations
 have practical implications to estimate beam intensities in rare-isotope beam facilities.

\acknowledgments
 This work was supported by  the
 National Key R$\&$D Program of China (Grant No.2023YFA1606403, 2023YFE0101500),
  the National Natural Science Foundation of China under Grants No.12475118, 12335007.
We also acknowledge the funding support from the State Key Laboratory of Nuclear Physics and Technology, Peking University (No. NPT2023ZX01).

% If you have acknowledgments, this puts in the proper section head.
%\begin{acknowledgments}
% put your acknowledgments here.
%\end{acknowledgments}

% Create the reference section using BibTeX:
% \bibliography{reference.bib}

\end{document}